\documentclass[10pt,conference]{IEEEtran}
\IEEEoverridecommandlockouts

\usepackage{array}  
\usepackage[colorinlistoftodos]{todonotes}
\usepackage{booktabs}
\usepackage{cite}
\usepackage{amsmath,amssymb,amsfonts}
\usepackage{algorithmic}
\usepackage{graphicx}
\usepackage{multirow}
\usepackage{siunitx}
\usepackage{textcomp}
\usepackage{xcolor}
\usepackage{placeins}
\usepackage{setspace}

\usepackage{threeparttable}
\usepackage[linesnumbered,ruled,vlined]{algorithm2e}

\def\BibTeX{{\rm B\kern-.05em{\sc i\kern-.025em b}\kern-.08em
    T\kern-.1667em\lower.7ex\hbox{E}\kern-.125emX}}
\begin{document}
\setstretch{0.97}

\title{{\LARGE Spec2Assertion: Automatic Pre-RTL Assertion Generation \\by LLMs with Progressive Regularization}

}
\author{\IEEEauthorblockN{Fenghua Wu, Evan Pan, Rahul Kande, Michael Quinn,\\
Aakash Tyagi, David Kebo Houngninou, Jeyavijayan (JV) Rajendran and Jiang Hu}
\IEEEauthorblockA{{Texas A\&M University}}}

\maketitle

\begin{abstract}
SystemVerilog Assertions (SVAs) play a critical role in detecting and debugging functional bugs in digital chip design. However, generating SVAs has traditionally been a manual, labor-intensive, and error-prone process. Recent advances in automatic assertion generation, particularly those using machine learning and large language models (LLMs), have shown promising potential, though most approaches remain in the early stages of development. In this work, we introduce Spec2Assertion, a new technique for automatically generating assertions from design specifications prior to RTL implementation. It leverages LLMs with progressive regularization and incorporates Chain-of-Thought (CoT) prompting to guide assertion synthesis. Additionally, we propose a new evaluation methodology that assesses assertion quality across a broad range of scenarios. Experiments on multiple benchmark designs show that Spec2Assertion generates $70\%$ more syntax-correct assertions with $2X$ quality improvement on average compared to a recent state-of-the-art approach. 

\end{abstract}


\section{Introduction}

As modern chip designs grow in complexity, ensuring their functional correctness becomes a critical challenge. Among the many approaches to functional verification, \textit{Assertion-Based Verification} (ABV) has long been an effective technique that facilitates fast bug detection and easier debugging via formally checking properties of expected behavior~\cite{witharanaSurveyAssertionbasedHardware2022, fosterAssertionBasedVerificationIndustry2008}. Assertions serve as localized monitors embedded in the design or testbench that can validate signal properties, control flow, timing constraints, and interface protocols~\cite{mehtaSystemVerilogAssertions2020}. They play a central role in simulation-based debug, coverage collection, and formal verification.

Despite their advantages, the process of crafting meaningful and correct assertions remains largely manual in practice. While assertions can be written in various forms and languages, one widely adopted standard in the hardware verification industry is the SystemVerilog Assertions (SVA)~\cite{fosterAssertionBasedVerificationIndustry2008}. Writing SVA requires deep understanding of the design intent, signal-level behavior, and specification semantics. This manual effort is error-prone, time-consuming, and requires domain expertise that is often scarce~\cite{fosterTrendsFunctionalVerification2015, fosterAssertionBasedVerificationIndustry2008}. 

To address this bottleneck, researchers have explored automatic assertion generation approaches. Early works~\cite{vasudevanGoldMineAutomaticAssertion2010,hertzMiningHardwareAssertions2013,palAssertionRankingUsing2020} apply static analysis and simulation trace mining to infer assertions, with GoldMine~\cite{vasudevanGoldMineAutomaticAssertion2010} being a representative seminal work.
Static analysis is a technique of examining RTL code without simulation, focusing on structural and behavioral properties such as data flow, control dependencies, and signal interactions. While practical and lightweight, these techniques are inherently constrained by simulation coverage~\cite{vasudevanGoldMineAutomaticAssertion2010} and cannot reason about unobserved or specification-driven behaviors. More structured approaches like HARM~\cite{germinianiHARMHintBasedAssertion2022} and A-TEAM~\cite{daneseATEAMAutomaticTemplatebased2017} rely on templates, hints, and pattern matching to mine assertions from RTL code. However, it is very difficult for limited templates or patterns to effectively deal with diversified designs. 

Traditional machine learning methods\cite{vasudevanGoldMineAutomaticAssertion2010,hertzMiningHardwareAssertions2013,ioannidesCoverageDirectedTestGeneration2012} can generate assertions from RTL code or simulation traces. However, they have limited capability in understanding natural language. Several other previous approaches~\cite{zhaoAutomaticAssertionGeneration2019,krishnamurthyEASEEnablingHardware2019} use natural language processing (NLP) to generate assertion by functional description. But these approaches need to manually extract the description from design specification and are therefore time consuming. In contrast, large language models (LLMs) like the GPT-4o series can directly comprehend complete natural language specification documents and extract complex semantic relationships without manual preprocessing. 

Recent advances have applied LLMs to the problem of assertion generation, aiming to bridge the gap between high-level design intent and formal properties. These efforts can be broadly categorized into two directions.
The first class of approaches employs LLMs to translate natural language specifications into assertions, without access to RTL code~\cite{fangAssertLLMGeneratingEvaluating2024,aditiHybridRulebasedMachine2022}. For example, AssertLLM\cite{fangAssertLLMGeneratingEvaluating2024} is a state-of-the-art technique that utilizes a multi-LLM prompting to extract key signals and behaviors directly from textual specifications. While this avoids dependency on RTL, it often leads to the generation of superficial or overly generic assertions. As observed in our experiments, such methods tend to produce a large number of structural checks—such as bitwidth validation or connectivity assertions—while failing to capture deeper functional properties that relate to the actual operation of the design. 

The second direction incorporates RTL design code into the assertion generation process, typically by concatenating RTL with spec text or analyzing signal names and structures within the code~\cite{sunImprovingVerificationProductivity,pulavarthiAreLLMsReady2025,baiAssertionForgeEnhancingFormal2025,kandeSecurityAssertionsLarge2024}. While this integration provides contextual grounding, it introduces a serious risk: the quality and correctness of generated assertions become tightly coupled with the RTL implementation. In practice, RTL descriptions—especially in early-stage or buggy designs—may contain errors, incomplete logic, or misleading signal behaviors. As a result, LLMs conditioned on such RTL may generate assertions that are syntactically correct but logically invalid or inconsistent with the intended specification. 

Overall, both the Spec+LLM and RTL+LLM approaches have their respective advantages and limitations.
We propose a new technique, called Spec2Assertion, within the Spec+LLM framework that aims to overcome the shortcomings of prior Spec+LLM methods while preserving their strengths. Specifically, our approach utilizes design specifications in a structured, RTL-independent manner, while ensuring formal soundness and supporting rich contextual reasoning.
These strengths are achieved through progressive regularization and Chain-of-Thought promting in using LLMs.
The main contributions of this work are summarized as follows.

\begin{itemize}

\item We propose a new function description extraction technique that can cover a wider scope than previous Spec+LLM techniques. 
\item To the best of our knowledge, Spec2Assertion is the first technique that leverages formal languages and Chain-of-Thoughts prompting in LLM-based assertion generation. 

    \item We develop a {\bf comprehensive technique for evaluating assertion quality}, addressing the limitations of prior works that either focus solely on quantity or assess quality only in a limited set of constrained scenarios.
    \item To the best of our knowledge, this work provides the first experimental comparison between LLM-based assertion generation and the seminal work of traditional ML approach GoldMine~\cite{vasudevanGoldMineAutomaticAssertion2010}.
\end{itemize}

We evaluate Spec2Assertion on multiple design benchmarks, and compare it with a seminal ML approach GoldMine~\cite{vasudevanGoldMineAutomaticAssertion2010,hertzMiningHardwareAssertions2013}, and a recent state-of-the-art Spec+LLM technique AssertLLM~\cite{fangAssertLLMGeneratingEvaluating2024}. 
The results show that Spec2Assertion generates $70\%$ and $3X$ more syntax-correct assertions than AssertLLM and GoldMine, respectively.
Moreover, Spec2Assertion achieves $2X$ and $1.5X$ assertion importance scores on average compared to AssertLLM and GoldMine, respectively.

\section{Background}

\subsection{Large Language Models for Formal Languages}
With the rapid advancement of LLMs in natural language  processing and task reasoning, they have demonstrated unprecedented potential in structured tasks such as mathematical reasoning and code generation. Recent studies, such as the ~\cite{zhouDontTrustVerify2024,luInterGPSInterpretableGeometry2021,liFormalLLMIntegratingFormal2024,merrillFormalLanguageTheory2021}, have shown that LLMs can not only generate step-by-step solutions to math problems but also automatically transform informal natural language mathematical questions into formal languages (e.g., theorem statements in Isabelle\cite{zhouDontTrustVerify2024}) and verify their internal consistency using automated theorem provers. This \textit{autoformalization} capability greatly enhances the reliability assessment of LLM outputs and lays a foundation for their deployment in formal language contexts.

Unlike natural languages, which are often ambiguous and context-dependent, formal languages\cite{salomaaFormalLanguages1987} are designed to be unambiguous and interpretable by both humans and machines. They play a critical role in domains like mathematics, programming, and hardware verification, where correctness hinges on well-defined semantics. This trend is particularly significant in the field of hardware verification, especially in ABV (Assertion-Based Verification), where SVAs (SystemVerilog Assertions) are widely used to formally describe and verify signal behaviors. Similar to mathematical theorems, SVA are essentially formal language expressions whose correctness relies on precise alignment with logical semantics and contextual constraints. Therefore, applying LLMs to assertion generation tasks can also be viewed as automatically formalizing natural language specifications into verifiable logical expressions. 

\subsection{Chain-of-Thought Prompting}
Chain-of-Thought (CoT) prompting\cite{weiChainofThoughtPromptingElicits2022,lingDeductiveVerificationChainofThought,liStructuredChainofThoughtPrompting2025} is a prompting technique in which LLMs are guided to generate intermediate reasoning steps before arriving at a final answer. Instead of producing a direct output, the model is encouraged to articulate a logical sequence of steps that mirrors human-like reasoning. This approach has proven especially effective in tasks that require multi-step deduction, such as arithmetic, symbolic manipulation, and program synthesis.

CoT has emerged as a powerful mechanism to enhance the reasoning capabilities of LLMs, particularly in domains requiring structured logic such as mathematics and code generation. Theoretically, CoT enables LLMs to simulate recursive computation by explicitly unfolding intermediate reasoning steps. The work of \cite{fengRevealingMysteryChain} demonstrates that while bounded-depth transformers struggle to directly solve arithmetic and code generation problems, the introduction of CoT allows constant-size autoregressive models to handle these tasks effectively through step-by-step derivations, significantly improving their expressive power.
These developments illustrate that CoT not only improves task performance, but also provides a bridge between natural language instructions and formal, verifiable representations. This is particularly relevant in formal hardware verification, where design intent must be faithfully translated into assertions expressed in SVA. 

\subsection{Static Analysis for Assertion Mining}
Static analysis is a foundational technique in hardware verification that enables the examination of RTL designs without requiring execution or simulation. It operates on the source code to analyze control-flow, data dependencies, and structural hierarchies within the design. In the context of assertion mining\cite{hertzMiningHardwareAssertions2013}, static analysis is typically used to identify internal functions and connections of a design. 

These methods often construct abstract representations such as variable dependency graphs or finite state models to trace how values propagate through a circuit. Key static features used in these approaches include signal assignment patterns, conditionals, always-block structures, and temporal correlations between control and data signals. 
\begin{figure*}[!ht]
    \centering
    \includegraphics[width=1\linewidth]{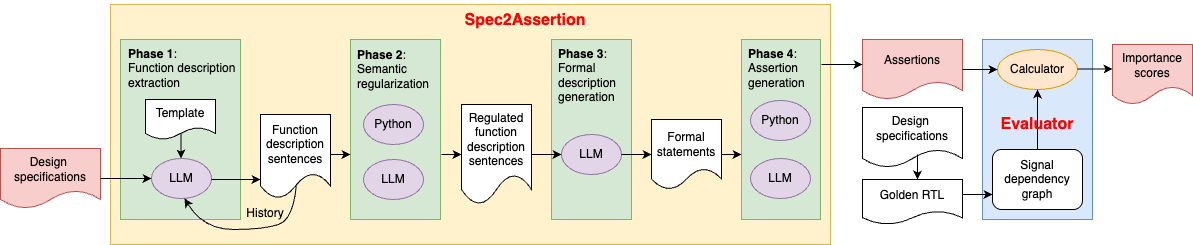}
    \caption{Overview of Spec2Assertion and its quality evaluation. }
    \label{fig:workflow}
\end{figure*}

\begin{figure*}[!ht]
    \centering
    \includegraphics[width=1\linewidth]{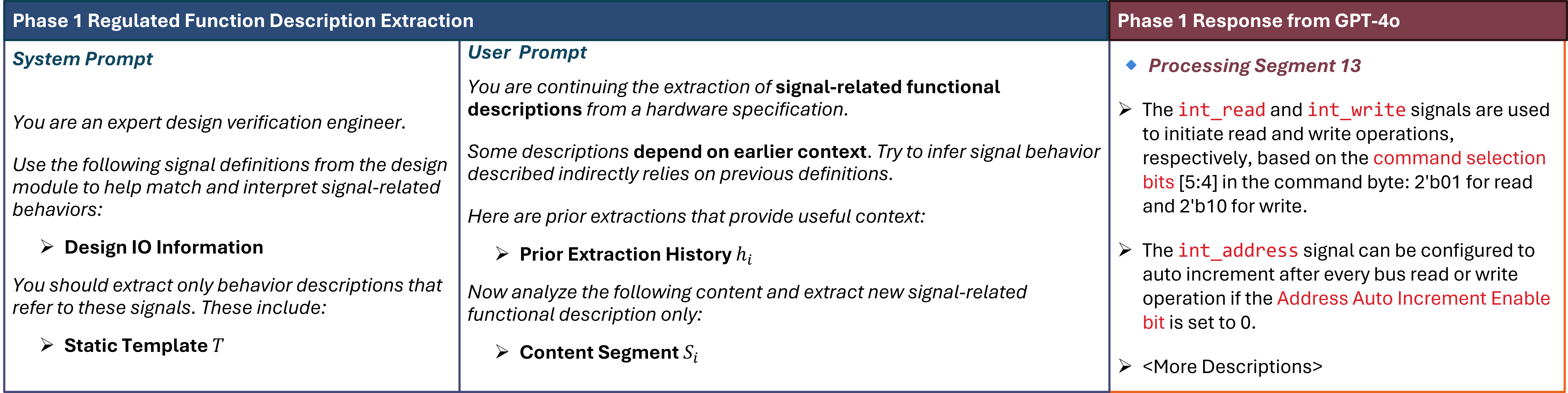}
    \caption{Phase 1. Regulated Function Description Extraction Prompt and the Response Example}
    \label{fig:Phase1Example}
\end{figure*}

Static analysis can also be used to evaluate the structural and temporal impact of assertions with respect to a design. This includes measuring signal connectivity, propagation depth, and temporal separation between antecedents and consequents (see Section~\ref{seq:assertionevaluation}). This allows us to maintain an RTL-agnostic generation process while leveraging structural analysis only for downstream quality evaluation.

\section{Spec2Assertion Methodology}
\label{sec:methodology}

\subsection{Overview}
\label{sec:overview}

Instead of directly generating assertions, which are formal and follow strict syntax, from design specifications that are free-form and informal, Spec2Assertion introduces a step-by-step regularization process as outlined in
Figure~\ref{fig:workflow}.
This progressive approach enables LLMs to transition smoothly through intermediate representations, making the task more manageable and effective compared to challenging one-shot or two-shot generation methods.
As such, Spec2Assertion Methodology is composed by four phases:
(1) regulated function description extraction,
(2) semantic regularization,
(3) formal description generation, and
(4) regulated assertion generation,
which are to be elaborated as follows.

\subsection{Phase 1: Regulated Function Description Extraction} 

This phase is to extract function descriptions from natural language design specifications. These descriptions are a crucial intermediate result that drives the next stages of assertion formalization and generation. 
This is where the game is won or lost\cite{witharanaSurveyAssertionbasedHardware2022}. 

Specification documents often describe the same functional feature across multiple, disjoint sections. Within each section, the language used may include ambiguous pronouns, incomplete phrasing, or context-dependent expressions, making it hard to extract a coherent and complete description of the intended behavior. As a result, prior extraction methods frequently yield outputs that are informal\cite{fangAssertLLMGeneratingEvaluating2024} or inconsistent\cite{aditiHybridRulebasedMachine2022}, especially when relying on single-pass or superficial analysis. This fragmentation and ambiguity hinder accurate identification of key behaviors, such as signal constraints, state transitions, and timing dependencies, which are critical for formal verification. Consequently, the extracted function descriptions themselves often inherit these flaws, resulting in artifacts that are difficult to formalize or directly verify.

In Phase 1 of Spec2Assertion, the input design specification file $S$ is divided into a set of segments $\{s_1, s_2, ..., s_n\}$, with each segment containing $m$ sentences The choice of $m$ is made to balance granularity and ensure that the total input remains within the LLM's context window.. The LLM processes one segment per iteration, keeping the task size manageable. To address limitations in previous approaches, each segment $s_i$ is preceded by a template hint $T$ and followed by a historical context 
$h_i$, forming the complete prompt for iteration $i$. This prompt guides the LLM to produce a function description item $d_i$.

The static template $T$ is responsible for identifying assertion-relevant functional behaviors described in natural language specifications. Usually, it is based on a list of signals described by the specification and consists of the following elements.
\begin{itemize}
    \item \textbf{Signal values}:
    \begin{itemize}
        \item \texttt{When \textless condition\textgreater, \textless signal\textgreater = \textless value\textgreater}
        \item e.g. \texttt{When ENABLE is high, READY = 1}
    \end{itemize}
    
    \item \textbf{Signal transitions}:
    \begin{itemize}
        \item \texttt{\textless signal\textgreater\ changes from \textless val1\textgreater\ to \textless val2\textgreater}
        \item \texttt{Only \textless condition\textgreater}
        \item e.g. \texttt{DATA\_VALID changes from 0 to 1 only when CLK\_EN is high}
    \end{itemize}

    \item \textbf{Reset behaviors}:
    \begin{itemize}
        \item \texttt{Reset clears \textless signal\textgreater}
        \item \texttt{After reset is deasserted, \textless signal\textgreater\ must transition to 1}
        \item e.g. \texttt{After RESETN is deasserted, INIT\_DONE must transition to 1}
    \end{itemize}
\end{itemize}

The history context $h_i$ is composed by the $k$ description items generated in the previous iterations, i.e., $h_i=\{d_{i-1}, d_{i-2}, ..., d_{i-k} \}$. This is to maintain the description consistency across multiple segments. Fig.~\ref{fig:Phase1Example} shows the prompt and the corresponding response.

The template hint along with historical context facilitate a structured list of function description items that are both \textit{hallucination-free} and \textit{consistent with the design’s signal interface}.

\subsection{Phase 2: Semantic Regularization}
\label{sec:semantic_regularization}

The result of Phase 1 is a set of natural language sentences $\Sigma=\{\sigma_1, \sigma_2, ...\}$. They still contain redundancy and ambiguity, which are to be removed in Phase 2. This phase consists of on preparation step and three regularization steps. 

\noindent
{\bf Preparation step.}
An LLM is used to extract signal names and signal descriptions from the design specification $S$. The correspondence between the signal names and their descriptions is established and a mapping table $M$, illustrated by Table~\ref{tab:mappingtable}, is obtained.
This mapping table required to perform the second regularization step. 

\begin{table}[!h]
\caption{Examples of signal names and their corresponding descriptions.}
\label{tab:mappingtable}
\centering
\setlength{\tabcolsep}{12pt}
\renewcommand{\arraystretch}{1.3}
\begin{tabular}{@{}ll@{}}
\toprule
\textbf{Signal name} & \textbf{Signal description} \\
\midrule
\texttt{bus\_wen} & when data should be written to the bus \\
\midrule
\texttt{bus\_wdt} & data to be written to the bus \\
\bottomrule
\end{tabular}
\end{table}

\noindent
{\bf Regularization steps}
\begin{itemize}
    \item \underline{Step 1: redundancy removal}. We use Python to detect if there are multiple identical sentences in $\Sigma$. If so, only one of them is kept and the others are removed. 
    \item \underline{Step 2: signal description to name conversion}. Using the mapping table from the preparation step, all signal descriptions are placed with their corresponding names using an LLM. This step greatly improves consistency and clarity, and facilitates smooth execution of subsequent phases. 
    \item \underline{Step 3: non-critical sentence pruning}. A sentence $\sigma_i \in \Sigma$ is critical if it describes a causal relationship between two events (or signal transitions), e.g., a transition of signal $x$ causes a transition of signal $y$. An LLM is applied to detect if a sentence is critical or not. If not, this sentence is pruned from $\Sigma$.
\end{itemize}
Through non-critical sentence pruning, Spec2Assertion prioritizes assertions that verify causal relationships among signals, while filtering out simple factual checks, such as verifying bitwidth consistency between design and specification. Although assertions are valuable for bug detection and debugging, they can introduce significant simulation overhead. Therefore, more assertions do not necessarily equate to better verification. Basic factual checks are better suited for lightweight sanity-check tools (e.g., Synopsys SpyGlass\cite{synopsys_spyglass_lint}), not formal assertions. In this context, Spec2Assertion focuses on maximizing the quality and efficiency of assertions, rather than their quantity.

\begin{figure}[h]
    \centering
    \includegraphics[width=\linewidth]{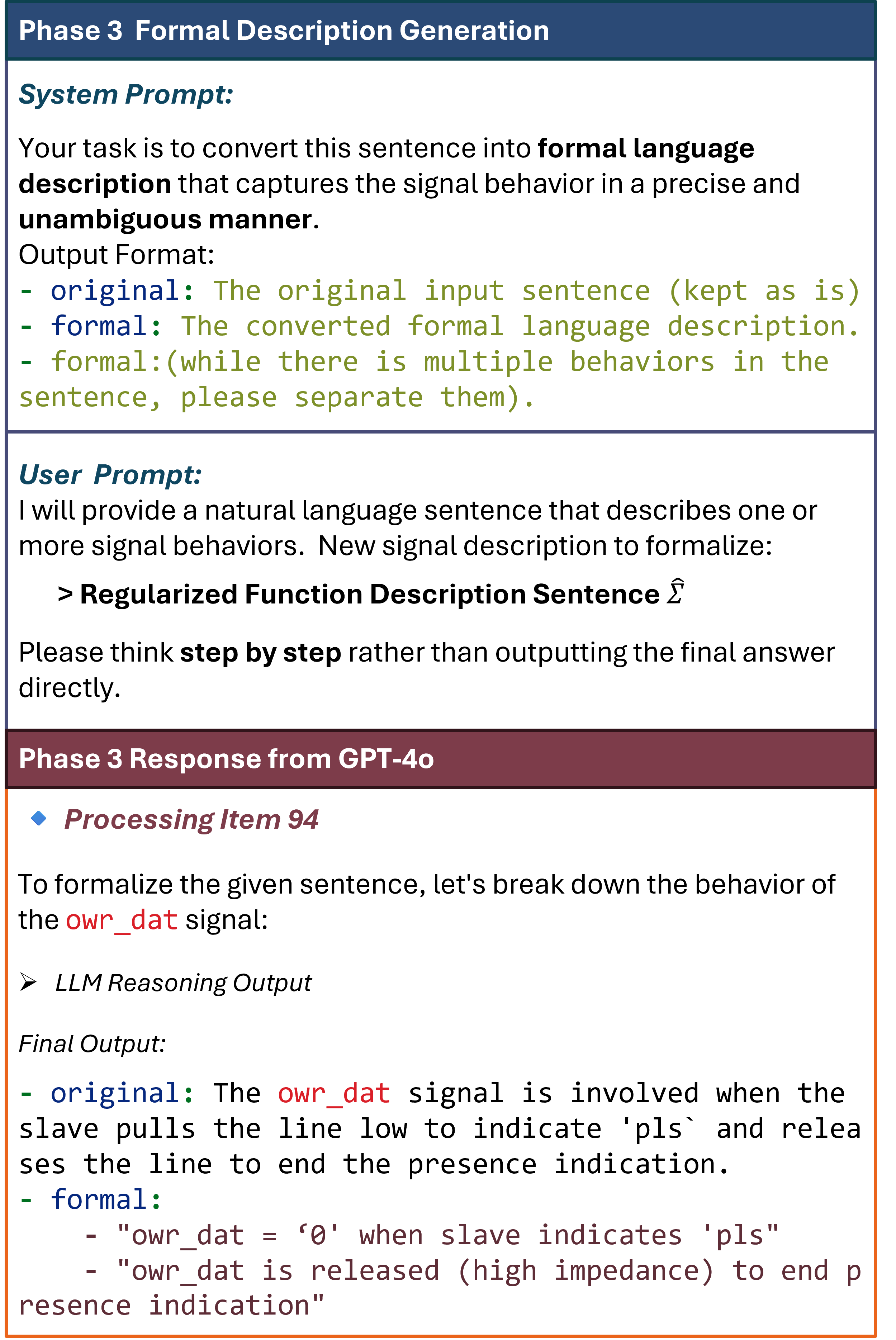}
    \caption{Phase 3. Formal Description Generation Prompt and Response Example}
    \label{fig:phase3}
\end{figure}
\subsection{Phase 3: Formal Description Generation}

Given the set of regularized function description sentences $\hat{\Sigma}$ from Phase 2, Phase 3 transforms each sentence into a formal language representation using an LLM as Fig.~\ref{fig:phase3}. This process iterates over $\hat{\Sigma}$, handling one sentence per iteration to maintain manageable context for the LLM. Popular LLMs, such as ChatGPT-4o~\cite{openaiGPT4TechnicalReport2024}, possess sufficient understanding of formal languages and can effectively translate natural language function descriptions into formal representations. Although various formal languages exist, Spec2Assertion does not depend on a specific one. Therefore, the LLM is guided to produce a formal language description without being constrained to a particular syntax.

A natural language sentence of function description sometimes mixes conditional triggers, temporal expectations, and functional effects. Consider the following description:
\paragraph{Example}
\begin{quote}
    \textit{``The grant policy for outport is first come first serve, if two inports request the same \texttt{tx\_outport\_req} simultaneously, the inport with the higher priority is favored.''}
\end{quote}

This sentence covers two verification objects:
\begin{enumerate}
    \item \textbf{First-come-first-serve:} \\
    Condition: An \texttt{inport} requests \texttt{tx\_outport\_req}. \\
    Action: Grant access to the first requesting \texttt{inport}.
    
    \item \textbf{Priority override:} \\
    Condition: Two \texttt{inports} request the same \texttt{tx\_outport\_req} simultaneously.\\
    Action: Grant access to the \texttt{inport} with the higher priority.
\end{enumerate}
The mixture of two verification objects may lead to incomplete coverage or missed semantics in resulting assertions,  especially when implicit causal, temporal, or functional behaviors are not explicitly separated beforehand.

To solve this challenge, we adopt Chain-of-Thoughts prompting so that the LLM can decompose multiple verification objects in a natural language function description sentence into multiple separated formal language statements. 

\subsection{Phase 4: Regulated Assertion Generation}

Given a set of formal language statements
$\Phi=\{\phi_1, \phi_2, ...\}$, Phase 4 generates SVAs. SVA is chosen for its industry-wide adoption, rich temporal operators, and compatibility with formal tools such as JasperGold~\cite{JasperAppsCommand2024}
This phase iterates through $\Phi$ and processes one statement in each iteration,
which consists of three steps.

\noindent
\underline{Step 1: Formal statement pruning}. A statement $\phi_i \in \Phi$ is pruned if it does not contain any signals in the mapping table $M$ obtained in Phase 2 (Section~\ref{sec:semantic_regularization}.This pruning is not performed in Phase 2 because the results of Phase 2 are function description sentences, each of which may lead to multiple formal statements in Phase 3. In other words, a function description sentence containing a signal in $M$ may become multiple formal statements, where some of the statement no longer contain the signal. This pruning is performed through Python. The pruning significantly reduces the likelihood of generating assertions that contain non-existent or irrelevant signals. 

\noindent
\underline{Step 2: Formal statement decomposition}. Each remaining formal statement $\phi$ is decomposed into a pair of antecedent $\alpha$ and consequent $\xi$ by the LLM. In logic theory, {\bf antecedent} is the condition that triggers a behavior and {\bf consequent} specifies the resulting behavior. The decomposition preserves the semantic fidelity of the specification, ensuring that the intended behavior is not distorted during the SVA generation process. In addition, the LLM is instructed to present the antecedents and consequents in the format of \texttt{sequence} constructs in SystemVerilog. This format facilitates easy realization of Step 3. 

\noindent    
\underline{Step 3: Assembling to SVAs}. 
We use Python to assemble each pair antecedent $\alpha$ and consequent $\xi$ into an SVA. Since both $\alpha$ and $\xi$ are already in SystemVerilog format, this step becomes straightforward.

\section{Evaluation of Automated Assertion Generation}
\label{seq:assertionevaluation}

\subsection{Motivation}
In evaluating the effectiveness of automatically generated assertions, prior LLM-based approaches~\cite{fangAssertLLMGeneratingEvaluating2024, baiAssertionForgeEnhancingFormal2025} have primarily focused on syntax correctness, which can be validated using software tools~\cite{synopsys_spyglass_lint,JasperAppsCommand2024}, and functional correctness, meaning the assertions behave as intended and can be verified using formal verification tools such as JasperGold~\cite{JasperAppsCommand2024}. However, these works largely overlook the quality of the generated assertions, an equally important dimension. For instance, assertions that check simple factual conditions (e.g., bitwidth matches) can be verified more efficiently using basic sanity check tools. Yet, they introduce nontrivial overhead to simulation runtime. In contrast, high-quality assertions are those that help expose and localize complex or subtle design bugs.
To address this gap, we propose augmenting assertion evaluation with a measure of assertion importance, which captures their relevance and utility in facilitating effective bug detection and root cause analysis.

\subsection{Prior Approach and Its Limitation}
An assertion importance metric was introduced in \cite{palAssertionRankingUsing2020} and used to evaluate assertions generated by GoldMine~\cite{vasudevanGoldMineAutomaticAssertion2010}. This metric is based on the PageRank algorithm~\cite{pagePageRankCitationRanking1999}, originally developed to rank webpages. However, the importance metric in \cite{palAssertionRankingUsing2020} is limited to cases where only a single signal appears in the consequent of an assertion, as is typical for assertions produced by GoldMine. As a result, it is not applicable when the consequent involves multiple signals.

\subsection{Signal Dependency Graph and Spatial Distance}
According to \cite{palAssertionRankingUsing2020}, static analysis is used to extract a Signal Dependency Graph (SDG) from an RTL design that has been previously verified to be functionally correct, referred to as the {\em golden RTL}. Please note that the golden RTL is used solely for evaluating the effectiveness of an assertion generation technique while the application of Spec2Assertion approach does not require access to RTL designs. Additionally, while the original work referred to this graph as a Variable Dependency Graph, we adopt the term Signal Dependency Graph to align with the terminology used throughout our work.

Each node in the Signal Dependency Graph (SDG) represents a signal transition (or event), and the edges capture the interdependencies between these events. Two key concepts are associated with SDGs:
\begin{itemize}
    \item \textbf{Spatial distance}:  
    Defined as the path depth between antecedent and consequent signals. A greater spatial distance indicates wider behavioral scope or stronger module-level interactions.   
    \item \textbf{Temporal distance}:  
    For multi-cycle behaviors, signal transitions at different clock cycles (e.g., \texttt{a@t}, \texttt{a@t+1}) are modeled as distinct temporal nodes in SDGs, capturing time-delayed causality.
\end{itemize}
\begin{figure}[ht]
  \centering
  \includegraphics[width=0.9\linewidth]{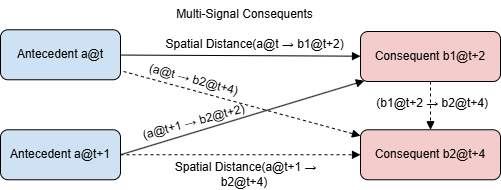}
  \caption{An example of signal dependency graph, where signal $a$ participates in two events in the antecedent, and there are two signals $b1$ and $b2$ with different transition times in the consequent. }
  \label{fig:multi-conseq-importance}
\end{figure}

An example of an SDG is illustrated in Figure~\ref{fig:multi-conseq-importance}. Spatial distance plays a central role in evaluating the importance of an assertion. When the spatial distance between an antecedent event and a consequent event is large, the corresponding bug becomes more difficult to detect, making the assertion more valuable. The significance of spatial distance has also been recognized in software assertion generation studies~\cite{ernst2003static}.

The concept of temporal distance involves unfolding a signal into multiple nodes within the SDG to represent different time points. For example, signal $a$ is unfolded into two separate nodes in Figure~\ref{fig:multi-conseq-importance}, each corresponding to a different clock cycle. Temporal distance also helps define ordering between multiple consequent events, such as $b1$ occurring before $b2$—as shown in the figure.

\subsection{Importance Score for Multi-Signal Consequent Assertions}

Based on \cite{palAssertionRankingUsing2020}, we propose an importance metric for general assertion cases including those with multiple signals in a consequent. 

Consider the following example assertion describing a handshake timeout protocol:

\begin{quote}
\small
\texttt{req \&\& !ack |-> \#\#3 timeout \#\#1 retry}
\end{quote}

This assertion specifies that:
\begin{itemize}
    \item If a {\em request} is issued but {\em no acknowledgment} is received,
    \item Then after three cycles, a {\em timeout flag} should be set,
    \item After another cycle, a {\em retry signal} should be triggered.
\end{itemize}

Our method evaluates such multi-signal assertions through two steps described as follows. The first step is temporal indexing,
where signals are temporally indexed (e.g., \texttt{req@t,ack@t}, \texttt{timeout\_flag@t+3}, \texttt{retry\_signal@t+4}) to represent time-shifted behavior within the SDG. Next, consider an assertion $\theta$ with multiple consequent events $C=\{c_1, c_2, ...,\}$. Each consequent event $c_i \in C$ has a set of causal events $P_i = \{p_{i,1}, p_{i,2}, ...\}$, where each event $p_{i,j} \in P_i$ is either an antecedent event that $c_i$ depends on or a consequent event in the same assertion that occurs before $c_i$, e.g., 
$b1$ occurs before $b2$ in Figure~\ref{fig:multi-conseq-importance}. Then, the {\bf importance score} for $\theta$ is defined as
\begin{equation}
    Importance(\theta) = \sum_{\forall c_i \in C} \sum_{\forall p_{i,j}\in P_i} D_{spatial}(p_{i,j}, c_i) \label{equ:importance}
\end{equation}

where $D_{spatial}(p_{i,j}, c_i)$ is the spatial distance between $p_{i,j}$ and $c_i$.

In order to model the recursive propagation of \textbf{spatial} distance, we define two directly connected signals \( \mathcal{A} \) and \( \mathcal{B} \) where \( \mathcal{B} \) depends on \( \mathcal{A} \). Specifically, for each direct dependency \( \mathcal{A} \to \mathcal{B} \), we have:
\begin{equation}
D_{spatial}(\mathcal{A}) = w(\mathcal{A}, \mathcal{B}) \times D_{spatial}(\mathcal{B}) + PR(\text{SDG}, \mathcal{A})
\label{eq:dspatial_local}
\end{equation}
where:
\begin{itemize}
    \item \( w(\mathcal{A},\mathcal{B}) \) denotes the number of times that \( \mathcal{A} \) directly influences node \( \mathcal{B} \) in the RTL.
    \item \( D_{spatial}(\mathcal{B}) \) aggregates importance recursively from \( \mathcal{B} \) and its direct dependency sources.
    \item \( PR(\text{SDG}, \mathcal{A}) \) represents the PageRank~\cite{pagePageRankCitationRanking1999} of \( \mathcal{A} \) within the SDG.
\end{itemize}

Through this recursive formulation, spatial distance propagates backward from consequent signals toward their dependent antecedents, combining local structural connectivity and global centrality.

\section{Experiment}

\subsection{Experiment Setup}

\subsubsection{Design Benchmarks}

We employ a suite of publicly available RTL designs from OpenCores \cite{opencoresOpenCoresOpenSource} and academic benchmarks, as listed in Table \ref{tab:design_benchmarks}. This is significantly more comprehensive than the recent work 
AssertLLM~\cite{fangAssertLLMGeneratingEvaluating2024}, where only one design is used in its evaluation. 
\begin{table}[h]
\renewcommand{\arraystretch}{1.2}
\setlength{\tabcolsep}{3pt}      
\centering
\footnotesize
\caption{Design benchmarks used for evaluation}
\begin{tabular}{|>{\raggedright\arraybackslash}m{2cm}|>{\raggedright\arraybackslash}m{2.2cm}|>{\raggedright\arraybackslash}m{4.1cm}|}
\hline
\textbf{Design} & \textbf{Source} & \textbf{Description} \\
\hline
\textbf{SOCKIT}       & OpenCores\cite{opencoresOpenCoresOpenSource} & Sockit controller on an FPGA-based SoC platform. \\
\hline
\textbf{UART}         & OpenCores\cite{opencoresOpenCoresOpenSource} & UART to bus interface. \\
\hline
\textbf{I2C}          & AssertLLM\cite{fangAssertLLMGeneratingEvaluating2024}         & I2C master/slave logic. \\
\hline
\textbf{HTAX}         & \cite{litzHyperTransportAdvancedXBar}         & Custom high-throughput arbitration and crossbar. \\
\hline
\end{tabular}
\label{tab:design_benchmarks}
\end{table}
These designs span various complexity levels and functional domains, facilitating a comprehensive evaluation. Each design has logic depth of 3-9 levels and 400-1200 lines of RTL code, which are all functionally correct. Please note the RTL designs here are for the sake of evaluation of assertion techniques while Spec2Assertion can produce assertions without RTL design. 

\subsubsection{Baseline Methods}
\begin{table*}[htbp]
  \centering
     \caption{Main comparison results.}
  \label{tab:evaluation_normalized}
  \small
  \renewcommand{\arraystretch}{1.5}
  \begin{threeparttable}
  \resizebox{1\textwidth}{!}{
  \begin{tabular}{l*{3}{c}*{3}{c}*{3}{c}*{3}{c}}
    \toprule
    & \multicolumn{3}{c}{\#non-trivial SVAs / \#SVAs}
    & \multicolumn{3}{c}{\#Syntax correct SVAs}
    & \multicolumn{3}{c}{\#SVAs passing formal verification}
    & \multicolumn{3}{c}{Average importance score} \\
    \cmidrule(lr){2-4} \cmidrule(lr){5-7} \cmidrule(lr){8-10} \cmidrule(lr){11-13}
    Benchmark
      & AssertLLM & GoldMine & \textbf{Spec2Assertion}
      & AssertLLM & GoldMine & \textbf{Spec2Assertion}
      & AssertLLM & GoldMine & \textbf{Spec2Assertion}
      & AssertLLM & GoldMine & \textbf{Spec2Assertion} \\
    \midrule
    UART
      & 48/59 & 0 & \textbf{54}/54
      & 42 & 0 & \textbf{47}
      & 9 & 0 & \textbf{17}
      & 0.12 & 0 & \textbf{0.15} \\
    HTAX
      & 73/90 & 41/83 & \textbf{153}/153
      & 17 & 41 & \textbf{152}
      & 2 & 2 & \textbf{15}
      & 0.00 & 0.14 & \textbf{0.17} \\
    I2C
      & 104/130 & 18/18 & \textbf{111}/111
      & 88 & 18 & \textbf{99}
      & 26 & 11 & \textbf{33}
      & 0.07 & 0.04 & \textbf{0.14} \\
    SOCKIT
      & 95/109 & 37/906 & \textbf{161}/161
      & 71 & 37 & \textbf{148}
      & 27 & 7 & \textbf{53}
      & 0.15 & \textbf{0.28} & 0.26 \\
     \midrule
    Normalized Avg.
      & 3.3X & 1X & 5X
      & 2.3X & 1X & 4X
      & 3.2X & 1X & 6X
      & 1X & 1.4X & 2.1X \\
    \bottomrule
  \end{tabular}
  }
  \end{threeparttable}
\end{table*}

In the experiment, Spec2Assertion is compared with the following two state-of-the-art techniques.

\begin{itemize}
    \item \textbf{AssertLLM}~\cite{fangAssertLLMGeneratingEvaluating2024}: 
    A recent LLM-based method that generates assertions directly from design specifications. Since its open-source document lacks critical details, it is reimplemented according to \cite{fangAssertLLMGeneratingEvaluating2024}.

    \item \textbf{GoldMine}~\cite{vasudevanGoldMineAutomaticAssertion2010}: 
    A seminal work of simulation- and mining-based tool that derives assertions from RTL traces. Its open-source code is used in generating results in this experiment. 
\end{itemize}
To the best of our knowledge, this is the first comparison between LLM-based assertion generation techniques and the seminal data mining approach GoldMine.

\subsubsection{The LLM}
GPT-4o~\cite{hurst2024gpt} is employed for both our Spec2Assertion and AssertLLM implementation. GPT-4o has a {\em temperature} parameter in range between 0 and 2 to tradeoff between exploitation and exploration. A high temperature implies more exploration, i.e., more creative solutions with a higher risk of errors. In our experiment, the default temperature is set to be 0.7.

\subsection{Results on Numbers of Generated SVAs}
Columns 2-4 in Table~\ref{tab:evaluation_normalized} show the numbers of SVAs (SystemVerilog Assertions) and numbers of non-trivial SVAs from these methods. AssertLLM~\cite{fangAssertLLMGeneratingEvaluating2024} generates significant number of trivial assertions, which are simple factual checks. These checks can be performed using sanity check software tools~\cite{synopsys_spyglass_lint} instead of being considered in assertions, which incur remarkable verification runtime overhead. GoldMine~\cite{vasudevanGoldMineAutomaticAssertion2010} does not produce simple factual check assertions. However, it produces redundant duplicated assertions, which are also trivial. Among the three methods, only our Spec2Assertion method produces no trivial assertions. 

Why does GoldMine produce duplicated assertions?
GoldMine’s strategy is to mine assertions by inspecting each bit of multi‑bit signals in RTL simulation traces.  For instance, for an 8‑bit bus it generates eight separate properties, one per bit. When multiple buses appear, it enumerates all combinations among them.  This approach leads to a combinatorial explosion in the total number of assertions, producing a significant number of duplicated assertions. For example, GoldMine produces 
900 assertions for 420 lines or RTL code in design SOCKIT. Moreover, GoldMine sometimes fails in extracting clean and well aligned simulation traces and consequently generates no assertions at all, as shown in the result of UART.

\subsection{Results on Syntax Correctness of the Assertions}
Syntax correctness of the generated assertions are verified by JasperGold~\cite{JasperAppsCommand2024} and the results are shown in columns 5-7 of Table~\ref{tab:evaluation_normalized}. One can see that our Spec2Assertion generates an average 4X syntax correct assertions compared to GoldMine. On average, 
it also increases the number of syntax correct assertions by $70\%$ compared to AssertLLM.

\subsection{Results on Formal Verification of Assertions}

The generated assertions are submitted to JasperGold for formal verification. A formal verification failure for an assertion $\theta$ indicates that the design specification may be incomplete with respect to $\theta$. For instance, if the specification lacks sufficient detail about input conditions, JasperGold might explore a behavior space $\Psi$ that is either beyond its practical analysis capability or leads to unrealistic scenarios, cases that would not occur in actual operation. Therefore, a formal verification failure does not necessarily invalidate the assertion and the assertion may still function correctly in RTL simulations.
That said, for a given design specification, a higher number of assertions passing formal verification is desirable, as it reflects better coverage within the well-defined behavior space of the specification. As shown in Columns 8–10 of Table~\ref{tab:evaluation_normalized}, Spec2Assertion produces much more assertions that pass formal verification compared to GoldMine and AssertLLM.

\subsection{Results on Assertion Quality}

The importance score according to Equation~\eqref{equ:importance} is calculated for all syntax correct assertions and the average results for all designs and methods are summarized in columns 11-13 of Table~\ref{tab:evaluation_normalized}. Our Spec2Assertion achieves the highest importance scores on the first three designs and is slightly behind GoldMine on SOCKIT. For the average importance score across all designs, ours is $1.5X$ and $2.1X$ of GoldMine and AssertLLM, respectively. 

\begin{figure}[!htb]
  \centering
  \includegraphics[width=0.85\linewidth]{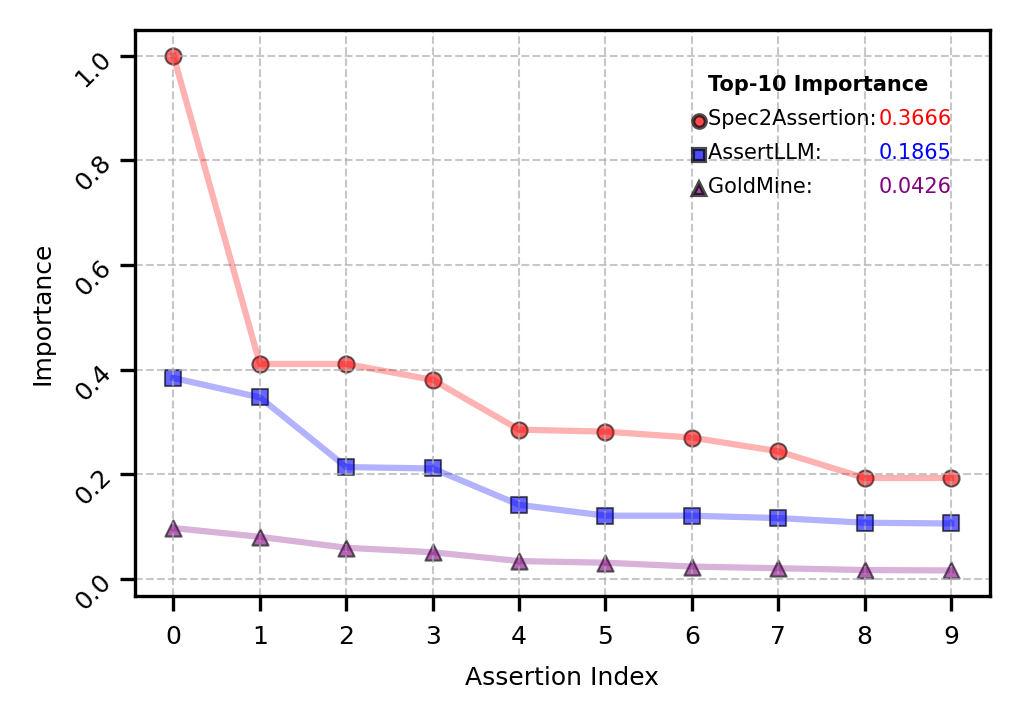}
  \caption{Top 10 normalized importance scores for I2C.}
  \label{fig:i2c_importance}
\end{figure}

\begin{figure}[!thb]
  \centering
  \includegraphics[width=0.85\linewidth]{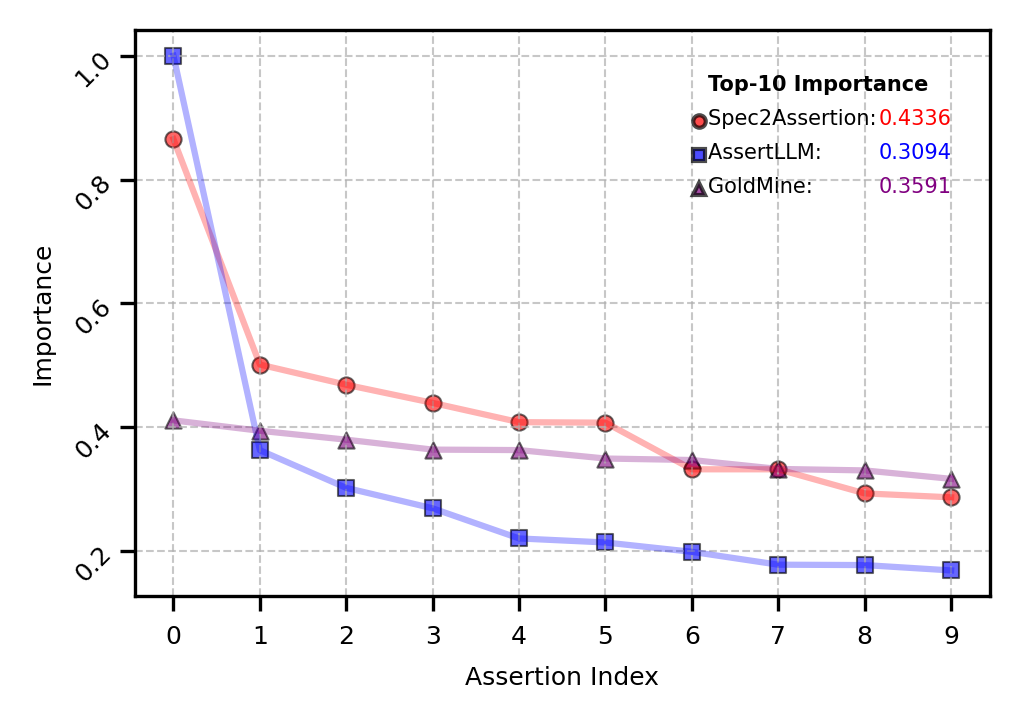}
  \caption{Top 10 normalized importance scores for SOCKIT.}
  \label{fig:sockit_importance}
\end{figure}

In Figure~\ref{fig:i2c_importance} and \ref{fig:sockit_importance}, we show importance scores of top 10 individual assertions from all methods for design I2C and SOCKIT, respectively. For I2C, our Spec2Assertion achieves importance scores consistently above the other two methods. 
For SOCKIT design, although AssertLLM briefly peaks the highest at assertion 0, the importance scores of its assertions quickly drops thereafter, whereas Spec2Assertion combines a strong initial score with a slower decrease, yielding the most semantically meaningful assertions throughout the top 10 assertions.

\subsection{Runtime Discussion}
Spec2Assertion takes approximately half an hour to generate assertions for each design. AssertLLM is slightly faster, while GoldMine requires about twice as much time due to its reliance on RTL simulations. Based on discussions with an industry engineer with decades of verification experience, manually generating the same set of assertions for each design would typically take several days. Overall, automatic assertion generation techniques offer roughly a two-order-of-magnitude speedup over manual efforts.

\subsection{Ablation Study}

To evaluate the contribution of the LLM parameter and key components in Spec2Assertion, we conduct an ablation study on the UART design, 

The main ablation study results are summarized in Table~\ref{tab:ablation_uart_small_norm} and the corresponding elaborations are provided as follows. 

\begin{table}[h]
\centering
\caption{Ablation results on UART design}
\label{tab:ablation_uart_small_norm}
\resizebox{0.5\textwidth}{!}{
\begin{tabular}{lcccc}
\toprule
\textbf{Setting} & \textbf{\#SVAs} & \textbf{\#Syn. correct} & \textbf{\#Proven} & \multicolumn{1}{c}{\textbf{Top10 imptc score}} \\
\midrule
Default (Temperature 0.7)    &  54 & 47 \textbf{(87\%)} & 17\textbf{ (31\%)} & \textbf{0.346}  \\
LLM Temperature 0.2   &  19 & 14 (74\%) &  3 (16\%) & 0.224  \\
No SemanticRegularization   & 180 & 75 (42\%) & 19 (11\%) & 0.230  \\
No FormalDescriptionGen.       &  49 & 39 (80\%) & 12 (24\%) & 0.195 \\
No Decomposition       &  54 & 36 (67\%) & 15 (28\%) & 0.020  \\
\bottomrule
\end{tabular}
}
\end{table}

\begin{itemize}
    \item \textbf{Lowering LLM temperature (0.2)}: By reducing the LLM’s  temperature from 0.7 to 0.2, only 19 assertions are generated, of which 14 are syntactically correct. The average importance score from the top 10 assertions is also dropped from 0.346 to 0.224.

    \item \textbf{Removing semantic regularization (Phase 2)}: Without semantic regularization, 180 assertions are generated, with 105 being syntactically correct and only 19 proven by formal verification. Although quantity increased, the quality and verifiability deteriorated. The average top-10 importance score decreases to 0.230, reflecting the generation of many unimportant assertions.

    \item \textbf{Removing formal description generation (Phase 3)}: Without formal description, the system directly passes raw specifications to the assertion generator. Under this setting, 49 assertions are generated, 39 are syntactically correct, and only 12 assertions are formally proven. This demonstrates that formal description improves both syntactic correctness and provability. The average top-10 importance score drops to 0.1954, indicating reduced relevance of generated assertions.

    \item \textbf{Removing statement decomposition (Phase 4)}: By skipping the decomposition step that modularizes formal statements into antecedent-consequent forms, 54 assertions are generated, but only 36 are syntactically correct, and 15 are proven. This demonstrates that decomposition significantly aids in improving syntax correctness and logical provability. The average top-10 importance score drastically falls to 0.0197, revealing that without decomposition, the critical assertions are largely missing.
\end{itemize}

\begin{figure}[h]
  \centering
  \includegraphics[width=0.85\linewidth]{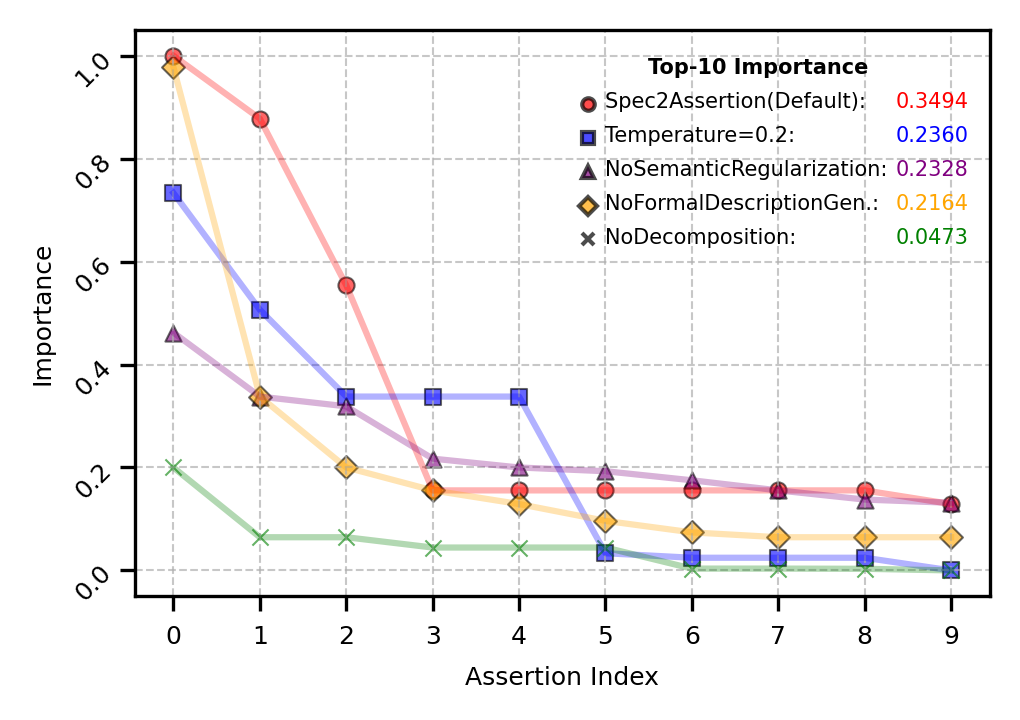}
  \caption{Normalized importance scores of the ablation study on UART.}
  \label{fig:uart_ablation}
\end{figure}

Individual normalized importance scores for the top 10 assertions in the ablation study are plotted in Figure~\ref{fig:uart_ablation}. The same trend as Table~\ref{tab:ablation_uart_small_norm} is observed and one can tell that the statement decomposition in Phase 4 is of critical importance. 

\section{Conclusion}
Spec2Assertion maintains high syntactic correctness and proven assertion rates, demonstrating not only generation capability but also practical verifiability. These results confirm the advantage of Spec2Assertion as the first known approach to apply signal-centric chaining and multi-LLM decomposition strategy in bridging specification intent with formal logic assertions.
Spec2Assertion distinguishes itself by uniquely leveraging design specifications, rather than relying on RTL implementations or simulation traces, as seen in other state-of-the-art approaches. By operating directly from specifications, Spec2Assertion avoids pitfalls associated with RTL-dependent methods, such as inaccuracies due to incomplete or erroneous RTL code. Further, the quality of assertions generated by Spec2Assertion is demonstrably superior with significantly higher importance scores, emphasizing the relevance and utility of these assertions in effective bug detection and root-cause analysis.

\bibliographystyle{IEEEtranS}
\bibliography{ref}

\begin{thebibliography}{10}
\providecommand{\url}[1]{#1}
\csname url@samestyle\endcsname
\providecommand{\newblock}{\relax}
\providecommand{\bibinfo}[2]{#2}
\providecommand{\BIBentrySTDinterwordspacing}{\spaceskip=0pt\relax}
\providecommand{\BIBentryALTinterwordstretchfactor}{4}
\providecommand{\BIBentryALTinterwordspacing}{\spaceskip=\fontdimen2\font plus
\BIBentryALTinterwordstretchfactor\fontdimen3\font minus \fontdimen4\font\relax}
\providecommand{\BIBforeignlanguage}[2]{{%
\expandafter\ifx\csname l@#1\endcsname\relax
\typeout{** WARNING: IEEEtranS.bst: No hyphenation pattern has been}%
\typeout{** loaded for the language `#1'. Using the pattern for}%
\typeout{** the default language instead.}%
\else
\language=\csname l@#1\endcsname
\fi
#2}}
\providecommand{\BIBdecl}{\relax}
\BIBdecl

\bibitem{aditiHybridRulebasedMachine2022}
F.~Aditi and M.~S. Hsiao, ``Hybrid {{Rule-based}} and {{Machine Learning System}} for {{Assertion Generation}} from {{Natural Language Specifications}},'' in \emph{2022 {{IEEE}} 31st {{Asian Test Symposium}} ({{ATS}})}.\hskip 1em plus 0.5em minus 0.4em\relax Taichung City, Taiwan: IEEE, Nov. 2022, pp. 126--131.

\bibitem{baiAssertionForgeEnhancingFormal2025}
Y.~Bai, G.~B. Hamad, S.~Suhaib, and H.~Ren, ``{{AssertionForge}}: {{Enhancing Formal Verification Assertion Generation}} with {{Structured Representation}} of {{Specifications}} and {{RTL}},'' Mar. 2025.

\bibitem{JasperAppsCommand2024}
Cadence, ``Jasper {{Apps Command Reference Manual}},'' 2024.

\bibitem{daneseATEAMAutomaticTemplatebased2017}
A.~Danese, N.~D. Riva, and G.~Pravadelli, ``A-{{TEAM}}: {{Automatic}} template-based assertion miner,'' in \emph{Proceedings of the 54th {{Annual Design Automation Conference}} 2017}.\hskip 1em plus 0.5em minus 0.4em\relax Austin TX USA: ACM, Jun. 2017, pp. 1--6.

\bibitem{ernst2003static}
M.~D. Ernst, ``Static and dynamic analysis: Synergy and duality,'' in \emph{WODA 2003: ICSE Workshop on Dynamic Analysis}, 2003, pp. 24--27.

\bibitem{fangAssertLLMGeneratingEvaluating2024}
W.~Fang, M.~Li, M.~Li, Z.~Yan, S.~Liu, H.~Zhang, and Z.~Xie, ``{{AssertLLM}}: {{Generating}} and {{Evaluating Hardware Verification Assertions}} from {{Design Specifications}} via {{Multi-LLMs}},'' Feb. 2024.

\bibitem{fengRevealingMysteryChain}
G.~Feng, B.~Zhang, Y.~Gu, H.~Ye, D.~He, and L.~Wang, ``Towards {{Revealing}} the {{Mystery}} behind {{Chain}} of {{Thought}}: {{A Theoretical Perspective}}.''

\bibitem{fosterAssertionBasedVerificationIndustry2008}
H.~Foster, ``Assertion-{{Based Verification}}: {{Industry Myths}} to {{Realities}} ({{Invited Tutorial}}),'' in \emph{Computer {{Aided Verification}}}, A.~Gupta and S.~Malik, Eds.\hskip 1em plus 0.5em minus 0.4em\relax Berlin, Heidelberg: Springer Berlin Heidelberg, 2008, pp. 5--10.

\bibitem{fosterTrendsFunctionalVerification2015}
H.~D. Foster, ``Trends in functional verification: A 2014 industry study,'' in \emph{Proceedings of the 52nd {{Annual Design Automation Conference}}}.\hskip 1em plus 0.5em minus 0.4em\relax San Francisco California: ACM, Jun. 2015, pp. 1--6.

\bibitem{germinianiHARMHintBasedAssertion2022}
S.~Germiniani and G.~Pravadelli, ``{{HARM}}: {{A Hint-Based Assertion Miner}},'' \emph{IEEE Transactions on Computer-Aided Design of Integrated Circuits and Systems}, vol.~41, no.~11, pp. 4277--4288, Nov. 2022.

\bibitem{hertzMiningHardwareAssertions2013}
S.~Hertz, D.~Sheridan, and S.~Vasudevan, ``Mining {{Hardware Assertions With Guidance From Static Analysis}},'' \emph{IEEE Transactions on Computer-Aided Design of Integrated Circuits and Systems}, vol.~32, no.~6, pp. 952--965, Jun. 2013.

\bibitem{hurst2024gpt}
A.~Hurst, A.~Lerer, A.~P. Goucher, A.~Perelman, A.~Ramesh, A.~Clark, A.~Ostrow, A.~Welihinda, A.~Hayes, A.~Radford \emph{et~al.}, ``Gpt-4o system card,'' \emph{arXiv preprint arXiv:2410.21276}, 2024.

\bibitem{ioannidesCoverageDirectedTestGeneration2012}
C.~Ioannides and K.~I. Eder, ``Coverage-{{Directed Test Generation Automated}} by {{Machine Learning}} -- {{A Review}},'' \emph{ACM Transactions on Design Automation of Electronic Systems}, vol.~17, no.~1, pp. 1--21, Jan. 2012.

\bibitem{kandeSecurityAssertionsLarge2024}
R.~Kande, H.~Pearce, B.~Tan, B.~{Dolan-Gavitt}, S.~Thakur, R.~Karri, and J.~Rajendran, ``({{Security}}) {{Assertions}} by {{Large Language Models}},'' \emph{IEEE Transactions on Information Forensics and Security}, vol.~19, pp. 4374--4389, 2024.

\bibitem{krishnamurthyEASEEnablingHardware2019}
R.~Krishnamurthy and M.~S. Hsiao, ``{{EASE}}: {{Enabling Hardware Assertion Synthesis}} from {{English}},'' in \emph{Rules and {{Reasoning}}}, P.~Fodor, M.~Montali, D.~Calvanese, and D.~Roman, Eds.\hskip 1em plus 0.5em minus 0.4em\relax Cham: Springer International Publishing, 2019, pp. 82--96.

\bibitem{liStructuredChainofThoughtPrompting2025}
J.~Li, G.~Li, Y.~Li, and Z.~Jin, ``Structured {{Chain-of-Thought Prompting}} for {{Code Generation}},'' \emph{ACM Transactions on Software Engineering and Methodology}, vol.~34, no.~2, pp. 1--23, Feb. 2025.

\bibitem{liFormalLLMIntegratingFormal2024}
Z.~Li, W.~Hua, H.~Wang, H.~Zhu, and Y.~Zhang, ``Formal-{{LLM}}: {{Integrating Formal Language}} and {{Natural Language}} for {{Controllable LLM-based Agents}},'' Aug. 2024.

\bibitem{lingDeductiveVerificationChainofThought}
Z.~Ling, Y.~Fang, X.~Li, Z.~Huang, M.~Lee, R.~Memisevic, and H.~Su, ``Deductive {{Verification}} of {{Chain-of-Thought Reasoning}}.''

\bibitem{litzHyperTransportAdvancedXBar}
H.~Litz, ``{{HyperTransport Advanced X-Bar}} ({{HTAX}}) {{Specification}}.''

\bibitem{luInterGPSInterpretableGeometry2021}
P.~Lu, R.~Gong, S.~Jiang, L.~Qiu, S.~Huang, X.~Liang, and S.-C. Zhu, ``Inter-{{GPS}}: {{Interpretable Geometry Problem Solving}} with {{Formal Language}} and {{Symbolic Reasoning}},'' Jul. 2021.

\bibitem{mehtaSystemVerilogAssertions2020}
A.~B. Mehta, \emph{System {{Verilog Assertions}} and {{Functional Coverage}}: {{Guide}} to {{Language}}, {{Methodology}} and {{Applications}}}.\hskip 1em plus 0.5em minus 0.4em\relax Cham: Springer International Publishing, 2020.

\bibitem{merrillFormalLanguageTheory2021}
W.~Merrill, ``Formal {{Language Theory Meets Modern NLP}},'' Jul. 2021.

\bibitem{openaiGPT4TechnicalReport2024}
OpenAI, ``{{GPT-4 Technical Report}},'' Mar. 2024.

\bibitem{opencoresOpenCoresOpenSource}
OpenCores, ``{{OpenCores}}: {{Open}} source hardware designs,'' https://opencores.org/.

\bibitem{pagePageRankCitationRanking1999}
L.~Page, S.~Brin, R.~Motwani, and T.~Winograd, ``The {{PageRank Citation Ranking}}: {{Bringing Order}} to the {{Web}}.'' Stanford InfoLab, Technical {{Report}} 1999-66, Nov. 1999.

\bibitem{palAssertionRankingUsing2020}
D.~Pal, S.~Offenberger, and S.~Vasudevan, ``Assertion {{Ranking Using RTL Source Code Analysis}},'' \emph{IEEE Transactions on Computer-Aided Design of Integrated Circuits and Systems}, vol.~39, no.~8, pp. 1711--1724, Aug. 2020.

\bibitem{pulavarthiAreLLMsReady2025}
V.~Pulavarthi, D.~Nandal, S.~Dan, and D.~Pal, ``Are {{LLMs Ready}} for {{Practical Adoption}} for {{Assertion Generation}}?'' Feb. 2025.

\bibitem{salomaaFormalLanguages1987}
A.~Salomaa, \emph{Formal Languages}.\hskip 1em plus 0.5em minus 0.4em\relax Academic Press Professional, Inc., 1987.

\bibitem{sunImprovingVerificationProductivity}
C.~Sun, C.~Hahn, and C.~Trippel, ``Towards {{Improving Verification Productivity}} with {{Circuit-Aware Translation}} of {{Natural Language}} to {{SystemVerilog Assertions}},'' vol.~1, no.~1.

\bibitem{synopsys_spyglass_lint}
\BIBentryALTinterwordspacing
{Synopsys}, ``{SpyGlass Lint - Static Verification Solution},'' 2024. [Online]. Available: \url{https://www.synopsys.com/verification/static-and-formal-verification/spyglass/spyglass-lint.html}
\BIBentrySTDinterwordspacing

\bibitem{vasudevanGoldMineAutomaticAssertion2010}
S.~Vasudevan, D.~Sheridan, S.~Patel, D.~Tcheng, B.~Tuohy, and D.~Johnson, ``{{GoldMine}}: {{Automatic}} assertion generation using data mining and static analysis,'' in \emph{2010 {{Design}}, {{Automation}} \& {{Test}} in {{Europe Conference}} \& {{Exhibition}} ({{DATE}} 2010)}.\hskip 1em plus 0.5em minus 0.4em\relax Dresden: IEEE, Mar. 2010, pp. 626--629.

\bibitem{weiChainofThoughtPromptingElicits2022}
J.~Wei, X.~Wang, D.~Schuurmans, M.~Bosma, b.~{ichter}, F.~Xia, E.~Chi, Q.~V. Le, and D.~Zhou, ``Chain-of-{{Thought Prompting Elicits Reasoning}} in {{Large Language Models}},'' in \emph{Advances in {{Neural Information Processing Systems}}}, S.~Koyejo, S.~Mohamed, A.~Agarwal, D.~Belgrave, K.~Cho, and A.~Oh, Eds., vol.~35.\hskip 1em plus 0.5em minus 0.4em\relax Curran Associates, Inc., 2022, pp. 24\,824--24\,837.

\bibitem{witharanaSurveyAssertionbasedHardware2022}
H.~Witharana, Y.~Lyu, S.~Charles, and P.~Mishra, ``A {{Survey}} on {{Assertion-based Hardware Verification}},'' \emph{ACM Computing Surveys}, vol.~54, no. 11s, pp. 1--33, Jan. 2022.

\bibitem{zhaoAutomaticAssertionGeneration2019}
J.~Zhao and I.~G. Harris, ``Automatic {{Assertion Generation}} from {{Natural Language Specifications Using Subtree Analysis}},'' in \emph{2019 {{Design}}, {{Automation}} \& {{Test}} in {{Europe Conference}} \& {{Exhibition}} ({{DATE}})}.\hskip 1em plus 0.5em minus 0.4em\relax Florence, Italy: IEEE, Mar. 2019, pp. 598--601.

\bibitem{zhouDontTrustVerify2024}
J.~P. Zhou, C.~Staats, W.~Li, C.~Szegedy, K.~Q. Weinberger, and Y.~Wu, ``Don't {{Trust}}: {{Verify}} -- {{Grounding LLM Quantitative Reasoning}} with {{Autoformalization}},'' Mar. 2024.

\end{thebibliography}

\end{document}